\documentclass[%
 reprint,
 amsmath,amssymb,
 aps,
]{revtex4-2}

\usepackage{graphicx}
\usepackage{dcolumn}
\usepackage{bm}
\usepackage{hyperref}
\usepackage{xcolor}
\usepackage{soul}

\begin{document}

\preprint{APS/123-QED}

\title{Structured beam controlled super-resolution in quantum dots via rapid adiabatic passage}

\author{Partha Das$^{1,}$}
 \email{partha.2015@iitg.ac.in}
\author{Samit Kumar Hazra$^{1,2,}$}
 \email{samit@phy.iitkgp.ac.in}
\author{Tarak Nath Dey$^{1,}$}
 \email{tarak.dey@iitg.ac.in}
\address{$^1$Department of Physics, Indian Institute of Technology Guwahati, Guwahati 781039, Assam, India}

\address{$^2$ Department of Physics, Indian Institute of Technology Kharagpur, Kharagpur 721302, West Bengal, India}







\begin{abstract}
We theoretically investigate rapid adiabatic passage (RAP)-based super-resolution microscopy in a two-level quantum dot (QD) system. The system consists of  QD interacting with the two structured beams, accompanied by chirping and a time delay. The central concept of this work lies in the stimulated emission depletion (STED) microscopy technique. To understand this physical mechanism behind super-resolved spot formation, we use the variational master equation (ME) for density matrix where radiative and non-radiative decays are incorporated. A suitably chosen spatiotemporal envelope of the structured beams enables the formation of a super-resolved image. We have also removed unwanted low-intensity circular rings around the spot using the Bessel-modulated truncated structured Laguerre Gaussian (LG) and super-Gaussian (SG) beams. We have studied the temperature variation of the current imaging technique. The numerical results confirm that at low pulse areas, exciton–phonon coupling distorts the image, while at higher pulse areas, exciton–phonon decoupling preserves image resolution. Hence, the proposed scheme may open up nanoscale imaging and bioimaging applications with QDs.

\end{abstract}

\maketitle


\section{\label{sec:level1}INTRODUCTION}

Conventional optics fail to resolve the spot size of an image beyond a value comparable to the probing light wavelength. Ernst Abbe first realized that the primary constraint of resolution imaging comes from diffraction \cite{intro1}. Later it was mathematically formulated using Fourier transform theory \cite{intro2}. Defeating the diffraction barrier has been the key to achieving high-resolution imaging. The use of super-resolution microscopy techniques can overcome the diffraction limit. In STED microscopy \cite{intro5,intro6}, excitation and depletion light beams with Gaussian and doughnut intensity profiles are used to illuminate the sample simultaneously. The excitation beam excites the fluorescent molecules to the bright state, and the depletion beam turns them back to the dark state by stimulated emission. As a result, the central zero intensity area of the doughnut beam fluorophores stays in a bright state and produces a tighter focused image. The transition probability between bright and dark states depends on the intensity of the laser beam. STED microscopy has opened up a new platform for imaging at the nanoscale in material science and medical biology \cite{intro7,intro8}. In the literature, various material systems have exhibited super-resolution phenomena, including fluorescently doped single nanocrystals \cite{r2_1}, optically active polymers \cite{r2_2}, nitrogen vacancy centers in diamond \cite{r2_3}, perovskites \cite{r2_4}, and biological systems \cite{r2_5}. The fundamental idea of the STED process is the same across all material systems, but the implementation differs depending on each system's unique material properties, dimensions, and environments. In particular, STED involves fluorescence depletion, which is an incoherent process. Several other contemporary super-resolution techniques, such as photoactivated localization microscopy (PALM) \cite{intro_change1} and stochastic optical reconstruction microscopy (STORM) \cite{intro_change2}, have also enabled optical imaging beyond the diffraction limit \cite{intro9}. Among the more recent developments, MINFLUX microscopy \cite{intro_change3} has demonstrated remarkable localization precision. Nevertheless, these approaches generally rely on nonlinear fluorescence depletion or stochastic switching of emitters, and therefore primarily involve incoherent optical responses.


In contrast, RAP based imaging offers an alternative route that exploits the coherent manipulation of quantum states to achieve efficient population transfer \cite{intro11}. It uses a time-advance strong  RAP positive chirping pulse that transfers the population from the ground to the excited state. A second time-delayed RAP pulse with negative chirping can de-excite the population from the excited state to the ground state. Hence, the intensity-dependent RAP  in a two-level system acts like a nearly ideal `on' and `off' switch, which qualifies the critical criterion of super-resolution microscopy. With the advent of short, intense pulse and pulse-shaping technology, selective population transfer has become efficient \cite{intro_new1}. Population transfer through a chirp pulse is much more effective than the pulse without chirping \cite{intro12,intro_new2}. Moreover, a frequency-swept pulse-induced excitation is immune from Rabi oscillation \cite{intro13,intro14}. Under the adiabatic condition, a strong Rabi frequency can transfer the population at the desired level without decay and decoherence inducing losses \cite{intro15}.

Semiconductor QDs allow precise control over their size, shape, and composition, enabling more tailored RAP implementation than atomic systems \cite{intro16}. The above mentioned scheme explores the effect of coherent control under decoherence in solid-state systems.  
Experiments on optically driven InGaAs/GaAs QDs found the intensity damping of Rabi rotation (RR) due to longitudinal acoustic (LA) phonons \cite{intro17,intro18}. These self-assembled QDs interact with the phonons, limiting the excitonic transition's coherence \cite{intro19,intro20,intro21,intro22,intro23,intro24,intro25}. Various theoretical approaches have been proposed for investigating phonon interaction's role in the excited state's coherent population distribution. This includes the ME using perturbative expansion of the exciton phonon coupling in Markovian \cite{intro17,intro18,intro22,intro26} and non-Markovian limit \cite{intro19,intro27,intro28}, numerical techniques with path integral method \cite{intro21}, and correlation expansion \cite{intro20,intro29,intro30}.\\
\indent  
In this work, we have studied the RAP-based super-resolution microscopy technique in a semiconductor quantum dot. The QDs are made of InGaAs and embedded in GaAs wetting layers. Due to their strong carrier confinement, energy levels become discrete like atoms. Here, we consider the QD an effective two-level system by considering large biexciton binding energy. This model does not account for Auger recombination in this system. In this process, an electron and a hole recombine within the QD, transferring energy to eject an extra electron or a hole from the QD without producing light \cite{r2_8}. To account for such Auger loss in this system, it is necessary to consider all possible electron-hole configurations in the QD. Consideration of all such configurations complicates the system and, therefore, significantly increases the numerical complexity when determining exciton dynamics. Similar to the STED technique, two oppositely chirped structured lights with super-Gaussian and doughnut intensity are coupled to the two QD levels. The system also incorporates phonon interaction with the QD due to lattice vibration associated with the environmental temperature. Therefore, investigating phonon-mediated dephasing is mandatory for image formation. To illustrate the effect of temperature on imaging, we study the variational ME. 
We also consider a truncated spatial envelope of the Bessel-modulated SG and Bessel-modulated LG beam to improve the spot size resolution by stopping residual ground state population excitation in the form of an additional unwanted ring. The system shows an excellent super-resolved image for the QD, which is beyond the capability of the conventional imaging system. Therefore, this scheme's tunable optical properties of QD have potential applications in sensors, drug delivery, biomedical imaging \cite{intro31,intro32}, quantum communication, and quantum information \cite{intro33}.\\
\indent
The paper is organized as follows. Section I contains a brief introduction to super-resolution and its application in the QD medium. Section II presents the level system and theoretical formalism considering the phonon contribution using the variational ME. In Section III, we discuss the numerical results regarding super-resolution image formation with various controlling parameters such as light intensity, and temperature by removing the unwanted low-intensity ring around the central bright spot image. Finally, in section IV, we give a conclusion of the work.

\section{THEORETICAL FORMULATION}

\begin{figure}[ht]
	\centering
	\includegraphics[width=0.7\linewidth]{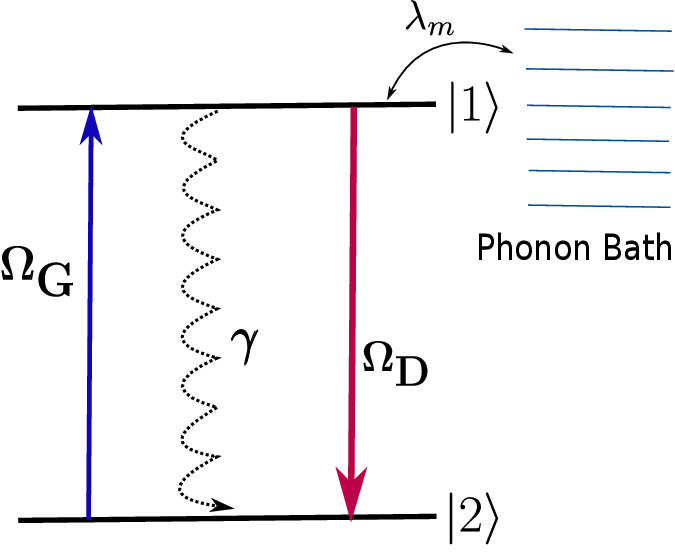}
	\caption{Schematic diagram of two-level quantum dot interacting with phonon bath. Two spatiotemporal beams interact with the system of Rabi frequencies $\Omega_{G}$ and $\Omega_{D}$.
		The spontaneous emission decay rate from $|1\rangle$ to $|2\rangle$ is given by $\gamma$. The two
		beams interact resonantly with both the levels.}
	\label{fig:1}
\end{figure}

Controllable population transfer at the excited state is the essence behind super-resolution imaging \cite{new3}. A variety of methods, such as stimulated Raman adiabatic passage (STIRAP) \cite{new4}, super-adiabatic STIRAP (saSTIRAP) \cite{new5}, and rapid adiabatic passage (RAP) \cite{intro12}, have been used to transfer the population to the desired state. The population inversion is beyond reach for a two-level system due to the thermodynamic limit. RAP can overcome this limitation \cite{new6}. The two-level system can achieve an efficient and robust time-dependent population inversion under the RAP. We study a detailed theoretical explanation for forming super-resolution imaging beyond the diffraction limit based on variational ME. The charge confinement of electron-hole pairs leads to semiconductor QD manifests an atom-like discrete energy level structure. A left-handed circularly polarized light that drives the excited (exciton) state $|1\rangle$ and ground state $|2\rangle$ with energy separation $\hbar\omega_{QD}$ produces a two-level configuration as shown in Fig. \ref{fig:1}.  The incident light consists of two spatiotemporal beams of opposite chirping interacting with the two-level quantum dot by the induced dipole moment.We have adopted a semi-classical treatment of light-matter interaction where the field is classical, and the energy levels of QD are discrete. The excitation and de-excitation of two beams which couple the states $| 1 \rangle$ and $| 2 \rangle$ are respectively given as
\begin{subequations}
\begin{align}
	&\Vec{E}_{G}(r,t) = \hat{\sigma}_- \mathcal{E}_{G}(r) e^{-\dfrac{t^2}{2\tau^2} - i \omega_{G} t -i\alpha_{G} \dfrac{t^2}{2}}+ \textit{c.c.}\label{eq:field1},\\
	&\Vec{E}_{D}(r,t) = \hat{\sigma}_- \mathcal{E}_{D}(r) e^{-\dfrac{t^2}{2\tau^2} - i \omega_{D} t -i\alpha_{D} \dfrac{t^2}{2} } + \textit{c.c.}\label{eq:field2},
\end{align}
\end{subequations}
where $\hat{\sigma}_-$ is the left circular polarization unit vector, $\tau$ is chirped pulse width, $\omega_{G}$ and $\omega_{D}$ are the corresponding carrier frequencies, $\alpha_{G}$ and $\alpha_{D}$ are the linear temporal chirp of the first and second pulse respectively. Specifically, the linear temporal chirp denotes the sweep rate of the laser from a negative detuning to a positive detuning or vice-versa. The spatial profiles of the two beams $E_G(r)$ and $E_D(r)$ are taken to be a SG and LG$_0^{1}$, respectively, expressed as 
\begin{subequations}
\begin{align}
	&\mathcal{E}_{G}(r) = \mathcal{E}_{G}^0 e^{-\left(\dfrac{r}{\sqrt{2} w_{G}}\right)^{2n} }, \\
	&\mathcal{E}_{D}(r) = \mathcal{E}_{D}^0 \left(\dfrac{r}{w_{D}}\right) e^{-\dfrac{r^2}{2 w_{D}^2}+im\phi}\label{eq:LG},
\end{align}
\end{subequations}
where $\mathcal{E}_{G}^0$ and $\mathcal{E}_{D}^0$ are the amplitude, $n$ is an integer which denotes the flatness of the Gaussian top. The value $n>1$ resembles the super-Gaussian or flat top beams and we have taken $n = 2$. The beam waist of each beam is denoted by $w_{G}$ and $w_{D}$, the orbital angular momentum index $l = 1$, and $\phi$ is the phase difference between the two beams, which we have taken $\pi/2$. The SG beam transfers the population from the ground state to the excited state, and the LG beam depletes the excited state population to the ground state. 

\subsection{System Hamiltonian}

The total  Hamiltonian of the system in the presence of the two beams under electric dipole approximation can be written as 
\begin{align}
	\textbf{\textit{H}} &=  \hbar \omega_{QD} |1 \rangle \langle 1 | -\Vec{d}_{12} . \left(\Vec{E}_{G}(r,t) + \Vec{E}_{D}(r,t) \right)|1 \rangle \langle 2 | \nonumber\\
	&+ \textit{H.c.},
\end{align}
where $\Vec{d}_{12} = \langle 1 | \hat{d} | 2 \rangle$ is the matrix elements of the induced dipole moment operator $\hat{d}$ for the transition $|1\rangle \leftrightarrow|2\rangle$. We make a unitary transformation as
\begin{align}
	&\mathbf{U} = e^{-i\nu(t) t |1\rangle \langle 1|},
\end{align}
where $\nu(t) = \omega_D + \alpha_D t/2$. Note that the time-dependent frequency signifies the characteristics of the chirp pulse. Now, the effective Hamiltonian in the interaction picture is given by
\begin{equation}
	\mathcal{H}_{eff} = \mathbf{U}^\dagger \textbf{\textit{H}} \mathbf{U} - i \hbar \mathbf{U}^\dagger \dfrac{\partial \mathbf{U}}{\partial t}.
\end{equation}
Under rotating wave approximation (RWA) the above Hamiltonian gives
\begin{align}
	\mathcal{H}_{eff} =& -\hbar \Delta(t) |1 \rangle \langle 1 | + \dfrac{\hbar}{2}(\Omega_G(r,t) e^{i \delta(t) t} + \Omega_D(r,t))|1 \rangle \langle 2 |\nonumber\\
	&+ \textit{H.c.},\nonumber
\end{align}
where the detunings are defined as
\begin{subequations}
	\label{eq:detuning}
	\begin{align}
		&\Delta(t) = (\omega_D + \alpha_D t) - \omega_{QD},\\
		&\delta(t) = (\omega_D - \omega_G) - \dfrac{t}{2}(\alpha_G  - \alpha_D ).
	\end{align}
\end{subequations}
Both the beams interact resonantly with two-level system $\textit{i.e.}$, $\omega_{D}=\omega_{G}=\omega_{QD}$. The spatiotemporal Rabi frequencies of respective beams are defined as, $\Omega_{G}(r,t) = \Vec{\mathbf{d}}_{12} . \hat{\sigma}_{-} \mathcal{E}_{G}e^{-t^2/2\tau^2}/\hbar$, and $\Omega_{D}^0 = \Vec{\mathbf{d}}_{12} . \hat{\sigma}_{-} \mathcal{E}_{D}e^{-t^2/2\tau^2}/\hbar$.

The system under consideration is significantly different from the well-studied single-atom emitters due to the solid-state nature of the semiconductor QD emitters. The medium consists of a few  InGaAs QDs grown on top of GaAs host material using molecular beam epitaxy. Therefore, the host lattice vibration modifies the QD dynamics depending on the environment temperature. In the literature, the quantized form of the vibrational energy in a periodic structure refers to a phonon. Many theoretical and experimental studies confirm the longitudinal acoustic (LA)  phonon coupling with QDs via deformation potential. Therefore, various new quantum phenomena were discovered, like the appearance of new features in Mollow triplets \cite{ep1,ep2}, emission line broadening \cite{intro20,ep3,ep4}, and limiting degree of indistinguishability of photons \cite{ep5,ep6}. On the other hand, the QD-phonon interaction model explains several quantum features, such as RR \cite{intro26}, RAP \cite{ep7,ep8,ep9}, and phonon-assisted state preparation \cite{ep10,ep11}. It is pertinent to include the effect of phonon, interacting with the two-level QD configuration. The system is coupled to an acoustic phonon bath represented by as a collection of harmonic oscillators with frequency $\omega_{m}$ = c$_s k$ where, c$_s$ is the velocity of sound and $k$ is the wavevector, creation and annihilation operator of the m$^{th}$ mode are $b_{m}^\dagger$ and $b_{m}$ respectively. The coupling constant for exciton phonon mode is $\lambda_{m}$. The effective Hamiltonian under the phonon consideration in the interaction picture can be written as \cite{ME1},
\begin{align}
	\textit{\textbf{H}}^{'}(t) &= - \hbar \Delta |1 \rangle \langle 1|+ \dfrac{\hbar }{2} (\Omega(t)|1 \rangle \langle 2| + \Omega(t)^*|2 \rangle \langle 1|) \nonumber\\
	&+ \sum_{m} \hbar \omega_m b_{m}^\dagger b_{m} + \sum_{m} \hbar \lambda_{m} (b_{m} + b_{m}^\dagger)|1 \rangle \langle 1|,	
\end{align}
where $\Delta$ is given by Eq. (\ref{eq:detuning}a) and the complex Rabi frequency $\Omega(t) = \Omega_D + \Omega_G e^{i\delta t}$. We make a variational transformation
\begin{equation}
	H_V= e^V H^{'} e^{-V},~ {\textrm {where}}~
	V = |1 \rangle \langle 1|\sum_{m} \dfrac{f_{m}}{\omega_{m}} (b_{m}^\dagger - b_{m} ).
\end{equation}
In the above equation the set of $f_m$ are the variational parameters. This transformed Hamiltonian gives the freedom to split the total Hamiltonian into system, bath, and interaction parts which are given as \cite{ME1,ME2},
\begin{subequations}
	\begin{align}
		&H_{SV} = -\hbar \Delta_v |1 \rangle \langle 1| + \langle B \rangle X_x(t) \\ 
		&H_B = \sum_{m} \hbar \omega_{m} b_{m}^\dagger b_{m} \\
		&H_{IV} = X_x(t)\zeta_x + X_y(t)\zeta_y + |1\rangle \langle 1| \zeta_z, \label{interaction_hamiltonian}
	\end{align}	
\end{subequations}
where $\Delta_v = \Delta + R$, and the shift $R = \sum_{m}^{}f_m(f_m - 2\lambda_{m})/\omega_m$ depends on the variational parameters. The phonon-modified system operators can be defined as follows:
\begin{align}
	X_x(t) &= \dfrac{\hbar}{2}(\Omega |1 \rangle \langle 2| + \Omega^* |2 \rangle \langle 1|)\nonumber\\
	X_y(t) &= \dfrac{i \hbar}{2}(\Omega |1 \rangle \langle 2| - \Omega^* |2 \rangle \langle 1|)\nonumber
\end{align}
The fluctuation operators induced by the bath are $\zeta_x = (B_+ + B_- + 2\langle B \rangle)/2$, $\zeta_y = (B_+ - B_-)/2i$, and $\zeta_z = \sum_{m}^{}(\lambda_m - f_m)(b_{m}^\dagger + b_{m} )$. The phonon displacement operator can be expressed as
\begin{equation}
	\label{bath operator}
	B_\pm = \text{exp}\left[\pm \sum \dfrac{f_m}{\omega_m} (b_m^\dagger - b_m)\right].
\end{equation}
The displacement operators contain summation over all the phonon modes. We need to find the variational parameter $f_k$, and in order to do so we use the free energy minimization \cite{ME2} and get the self-consistent form of the free parameter as
\begin{equation} 
	\label{variational_parameters}
	f_m=\dfrac{\lambda_m\left[1-\dfrac{\Delta_v}{\eta_v}\text{tanh}(\hbar\beta\eta_v/2)\right]}{1-\dfrac{\Delta_v}{\eta_v}\text{tanh}(\hbar\beta\eta_v/2)\left[1-\dfrac{\Omega_v^2}{2\Delta_v \omega_m}\text{coth}(\hbar\beta\omega_m/2)\right]} ,
\end{equation}
where the inverse temperature $\beta = 1/k_BT$, $\Omega_v = \langle B \rangle |\Omega|$, and $\eta_v=\sqrt{\Omega_v^2 + \Delta_v^2}$. We note from Eq. (\ref{bath operator}) that the bath operators are now function of the variational parameters. We can average it out for a particular temperature $T$ as, $\langle B_+  \rangle = \langle B_- \rangle \equiv \langle B \rangle $. The expectation value $\langle B \rangle$ and the shift $R$ is given by
\begin{subequations}
	\label{bath_expectation}
\begin{align}	
\langle B \rangle &= \text{exp}\left[-\dfrac{1}{2} \int_0^\infty \dfrac{J(\omega)F(\omega)^2}{\omega^2}\text{coth} \left(\dfrac{\hbar\beta \omega}{2}\right) d\omega \right], \\
R &= \int_{0}^{\infty} \dfrac{J(\omega)F(\omega)}{\omega}\left[F(\omega) - 2\right] d\omega,
\end{align} 
\end{subequations}
where $F(\omega_m) = f_m/\lambda_{m}$, and for the deformation potential coupling of the exciton-phonon we can take the super-Ohmic spectral density $J(\omega) = \alpha_p \omega^3 \text{exp}\left[-\omega^2/\omega_b^2\right] $, where $\alpha_p$ and $\omega_b$ are the electron-phonon coupling strength and the phonon cutoff frequency, respectively. The set of Eqs. (\ref{bath_expectation}) is solved numerically in a self-consistent manner.

It is now instructive to examine the behavior of $f_m$ in two limiting cases from Eq. (\ref{variational_parameters}): (i) In the limit $|\Omega|\ll \omega_m$, we find $f_m \approx \lambda_{m}$, corresponding to the full polaron transformation \cite{ME3}. Here, the driving field is sufficiently weak that the bath oscillators can adiabatically follow the excitonic motion, becoming fully displaced when the system occupies its excited state, as dictated by the coupling term in $H'$. (ii) Conversely, when $|\Omega|\gg \omega_m$, the variational parameter $f_m$ becomes negligibly small, indicating that the transformation induces little to no displacement. In this regime, the excitonic dynamics are too rapid for the relevant phonon modes to respond, and their displacements are therefore strongly suppressed. Later we can observe that this distinction has significant physical implications for driven quantum dots at high driving strengths. The inability of the phonon environment to follow the system dynamics leads to a reduction in phonon-induced damping. This key feature can be captured naturally within the variational framework.

\subsection{Variational master equation}

To study the ME in the variational frame \cite{ME1}, we must first evaluate the relevant correlation functions. From the interaction Hamiltonian Eq. (\ref{interaction_hamiltonian}), it becomes evident that the variational transformation introduces two distinct types of contributions, one resembling that of the polaron theory, and the other similar to the weak-coupling approach. As a result, the final ME naturally consists of three components: weak-coupling–like terms, polaron-like terms, and cross terms that originate from the interaction between the two kinds of bath operators. In the limiting cases, it smoothly reduces to either the weak-coupling or the polaron description. However, under general conditions, both types of contributions coexist. The cross terms play a crucial role, enabling the theory to interpolate seamlessly between the weak-coupling and polaron regimes. Thus it captures a broader range of physical behavior than either approach alone. We can now obtain the variational ME including the radiative and dephasing rates as
\begin{equation}\label{eq:variational master equation}
	\dfrac{\partial \rho}{\partial t} = -\dfrac{i}{\hbar}[H_{SV}, \rho] + \dfrac{\gamma}{2}\mathcal{L}[\sigma^-] \rho + \dfrac{\gamma^\prime}{2}\mathcal{L}[\sigma^{+}\sigma^-] \rho + \mathcal{L}_{ph} \rho
\end{equation}
where $\sigma^+ = |1\rangle \langle 2|$, $\sigma^- = |2\rangle \langle 1|$ are raising and lowering operators of the system respectively. The Lindblad superoperator, $\mathcal{L}[\hat{O}]\rho = 2\hat{O}\rho\hat{O}^\dagger - \hat{O}^\dagger\hat{O}\rho - \rho\hat{O}^\dagger\hat{O}$ acting on a operator $\hat{O}$. The radiative decay and dephasing are denoted by $\gamma$ and $\gamma^\prime$, respectively. The term $\mathcal{L}_{ph}$ includes phonon bath in system dynamics is given by,
\begin{align}\label{eq:lph}
	\mathcal{L}_{ph}\rho =& -\dfrac{1}{\hbar^2} \int_0^\infty \sum_{ij} \{C_{ij}(\tau)[X_i(t),X_j(t,\tau)\rho(t)]\nonumber\\& + H.c.\}\ d\tau
\end{align}
where $i, j \in x,y,z$, $X_j(t,\tau) = e^{- i H_{SV} \tau / \hbar}X_j (t)e^{ i H_{SV} \tau / \hbar}$ and the correlation functions are,
\begin{subequations}
	\label{polaron_correlation}
	\begin{align}
		&C_{xx}(\tau) = \langle B \rangle^2 \{\cosh[\phi(\tau)] - 1\},\\
		&C_{yy}(\tau) = \langle B \rangle^2 \sinh[\phi(\tau)],   
	\end{align}
\end{subequations} 
and $C_{xy}(\tau)=C_{yx}(\tau) = 0$. The dependence on the phonon propagator is given by
\begin{subequations}
	\begin{align}
		\phi(\tau) =& \int_0^\infty \dfrac{J(\omega)F(\omega)^2}{\omega^2}\biggl[\coth \left(\dfrac{\beta\hbar \omega}{2}\right) \cos(\omega \tau)\nonumber\\ &-i \sin(\omega \tau)\biggl]d\omega.
	\end{align}
\end{subequations}
It depends on the variational parameters through the term $F(\omega)$. We can also find the weak-coupling and cross-coupling correlation functions $C_{zz}(\tau)$, and $C_{yz}(\tau)$ as \cite{ME1}
\begin{subequations}
	\label{weak_cross_correlation}
	\begin{align}
		C_{zz}(\tau) =& \int_0^\infty J(\omega)\left[1 - F(\omega)\right]^2\biggl[\coth \left(\dfrac{\beta\hbar \omega}{2 }\right) \cos(\omega \tau)\nonumber\\ &-i \sin(\omega \tau)\biggl]d\omega, \\
		C_{yz}(\tau) =& -\langle B \rangle \int_0^\infty \dfrac{J(\omega)F(\omega)}{\omega}\left[1 - F(\omega)\right]\nonumber\\ & \times\biggl[\coth \left(\dfrac{\beta\hbar \omega}{2}\right) \sin(\omega \tau)+i \cos(\omega \tau)\biggl]d\omega,
	\end{align}
\end{subequations}
and also $C_{zy}(\tau) = - C_{yz}(\tau)$, $C_{zx}(\tau) = C_{xz}(\tau) = 0$.
Now we can find all the phonon induced decay rates by using Eq. (\ref{eq:lph}) shown in the Appendix.

\subsection{Rapid adiabatic passage in two-level system}
Besides spectroscopy, it is of profound importance to prepare a specific quantum state in semiconductors for different fields like quantum computation \cite{rap1,rap2}, single and entangled photons \cite{rap3,rap4}, Bose-Einstein condensation\cite{rap5}. RAP is an advantageous way to prepare a state as it remains insensitive to the variation of the laser field intensity or the pulse area beyond the adiabatic threshold. The population change between the two-level atomic states can happen in two distinct adiabatic ways. To understand the two processes, we use the well-known dressed state (adiabatic state) eigenvectors of a time-dependent Hamiltonian of a two-level system interacting with a chirp pulse of Rabi frequency $\Omega_1$ is given as \cite{rap6}
\begin{subequations}
\begin{align}
	|\psi_+ (t)\rangle =& \sin\theta(t)~|2\rangle + \cos\theta(t)~|1\rangle, \\
	|\psi_- (t)\rangle =& \cos\theta(t)~|2\rangle - \sin\theta(t)~|1\rangle, 
\end{align}
\end{subequations}
where the instantaneous eigenstates are the linear superposition of the bare states (diabatic state). The mixing angle $\theta(t)$ between the states is defined by
\begin{subequations}
\begin{align}
	\sin 2\theta =& \dfrac{|\Omega_1(t)|}{\sqrt{\Delta_1(t)^2 + |\Omega_1(t)|^2}}, \\
	\cos 2\theta =& \dfrac{\Delta_1(t)}{\sqrt{\Delta_1(t)^2 + |\Omega_1(t)|^2}},  
\end{align}
\end{subequations}
where $\Delta_1(t)$ is the time-dependent detuning. We can also, obtain the eigenvalues of the dressed state which are given by
\begin{equation}{\label{eq.eigenevalues}}
	E_{\pm} = \dfrac{\hbar}{2}\left[ \Delta_1(t) \pm \sqrt{\Delta_1(t)^2 + |\Omega_1(t)|^2}\right].
\end{equation}
Indeed, we can write the Hamiltonian in the adiabatic basis as
\begin{equation}
	H_a = \hbar
	\left[\begin{matrix}
		E_- & -i\Dot{\theta} \\
		i\Dot{\theta} & E_+ \\
	\end{matrix}\right].
\end{equation}
The adiabatic condition requires that the maximum rate of change in the adiabatic states, $|\psi_{\pm}\rangle$ must be smaller than the minimum difference between the eigenvalues \cite{rap6}. Thus the adiabatic condition reads
\begin{equation}
|\Dot{\theta}(t)| \ll |E_+(t) - E_-(t)|.
\end{equation}
Now for the case of (i) constant detuning, we can infer from Eq. (\ref{eq.eigenevalues}) that the energies of the dressed states remain parallel to each other long before and after the interaction with a pulsed laser. Only during the pulse interaction time, the states are in the superposition of the bare states. Therefore, in this case, the population completely return to its initial state, which is the no-crossing scenario of adiabatic evolution. In another case of (ii) time-dependent detuning, $\textit{i.e.}$, when the frequency sweeps adiabatically from a large negative value to a large positive value (or vice versa) two limits arise, (a) for large negative detuning ($|\Omega_1(t)| \ll |\Delta_1(t)|$),
\begin{align}\label{eq:rap1}
	&E_+ \rightarrow 0 ;\hspace{0.3cm} E_- \rightarrow -\hbar\Delta_1,\nonumber\\
	&|\psi_+\rangle \rightarrow |2\rangle ;\hspace{0.3cm} |\psi_-\rangle \rightarrow -|1\rangle,
\end{align}
and (b) for large positive detuning ($|\Omega_1(t)| \ll |\Delta_1(t)|$),
\begin{align}\label{eq:rap2}
	&E_+ \rightarrow \hbar\Delta_1 ;\hspace{0.3cm} E_- \rightarrow 0, \nonumber\\
	&|\psi_+\rangle \rightarrow |1\rangle ;\hspace{0.3cm} |\psi_-\rangle \rightarrow |2\rangle.
\end{align}
Both the two limits assert that the initial population in state $|2\rangle$ adiabatically follows $|\psi_+\rangle$ during the frequency-swept and finally makes an inversion to state $|1\rangle$. This is called the avoided crossing or anticrossing in adiabatic evolution. It is also called rapid as the process should occur shorter than the lifetime of the excited state.

\section{Numerical Results}
\subsection{RAP-based spot formation}
Under the adiabatic condition, a two-level system interacting with a Gaussian chirp pulse can robustly transfer the population from one state to another. We 
\begin{figure}[ht]
	\centering
	\includegraphics[width=\linewidth]{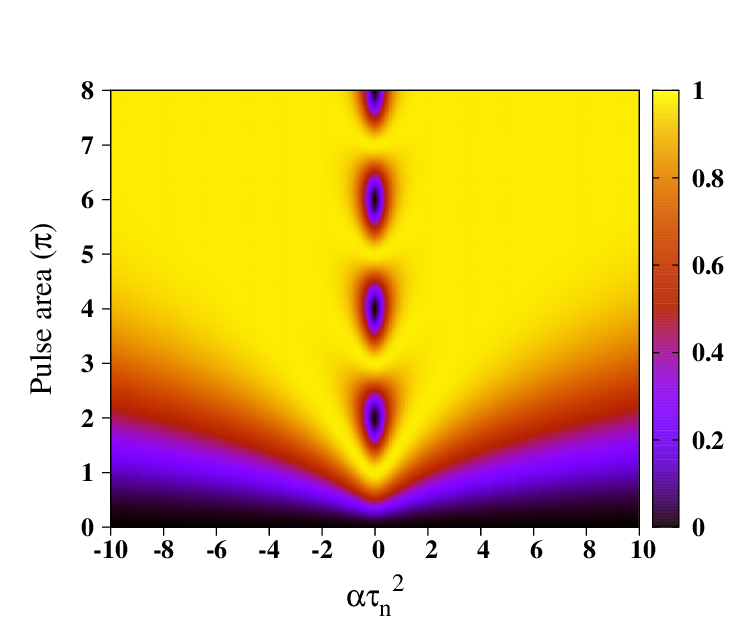}
	\caption{Excited state population as a function of pulse area and chirping. The color bar denotes the population of the excited state.}
	\label{fig:2}
\end{figure}
have used two such sequential Gaussian chirp pulses with opposite chirping. 
\begin{figure}[ht]
	\centering
	\includegraphics[width=\linewidth]{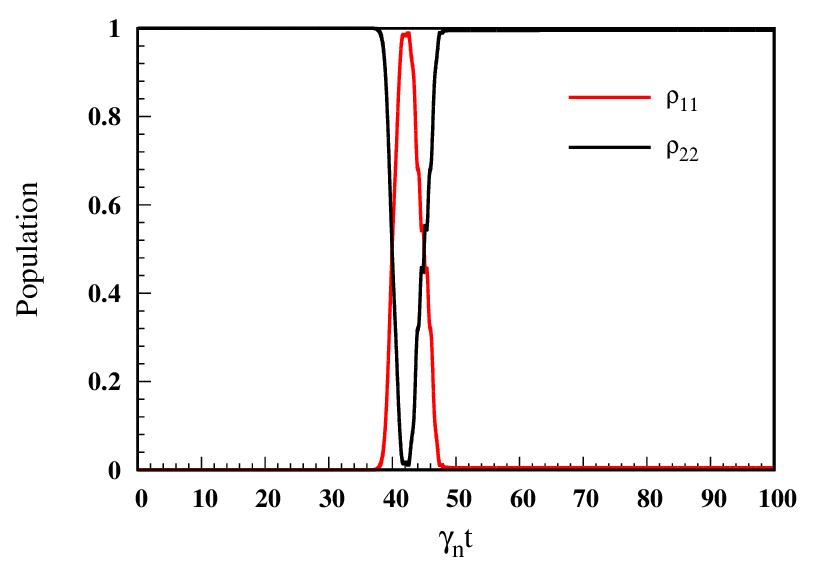}
	\caption{Population transfer via RAP. The first pulse with positive chirp of 3.24 ps$^{-2}$ and  pulse area greater than $\pi$ transfers the population to the excited state and the subsequent pulse with chirp -3.24 ps$^{-2}$ and  pulse area greater than $\pi$ bring down the excited state population to ground state. Both the pulse has a width 1.3$\tau_n$. The electron-phonon coupling strength $\alpha_p$ = 0.027 ps$^2$. }
	\label{fig:3}
\end{figure}
To perform the computation with the system parameters, we need to normalize it to a dimensionless quantity. We have chosen the normalized frequency $\gamma_{n}$ = 1 rad ps$^{-1}$ and time $\tau_n = 1/\gamma_n$. In Fig. \ref{fig:2}, we have plotted the excited state population as a function of the pulse area and chirp ($\alpha$) of the pulse without taking the interaction with phonon bath. We find if there is no chirp ($\alpha\tau_n^2$ = 0), we get the usual Rabi oscillation $\textit{i.e.}$, for the odd multiple of $\pi$ the population transfer to the excited state whereas for even multiple of $\pi$ it remains in the ground state. Also, we notice that robust population transfer can happen due to RAP for several parameters that satisfy the conditions, $|\alpha|\tau_n^2>>1$ and $|\alpha|\tau_n^2 << \Omega_0^2\tau_n^2$. Here $\Omega_0$ is the peak Rabi frequency of the chirped pulse. As in Fig. \ref{fig:3}, the first pulse peaked at $\gamma_nt$ = 10, Rabi frequency $\Omega_{G}^0 = 4.0\gamma_{n}$, positive chirp $\alpha_G$ = 3.24 ps$^{-2}$ and having pulse area $>\pi$ takes the population to the excited state that was initially in the ground state. 
\begin{figure}[ht]
	\centering
	\includegraphics[width=\linewidth]{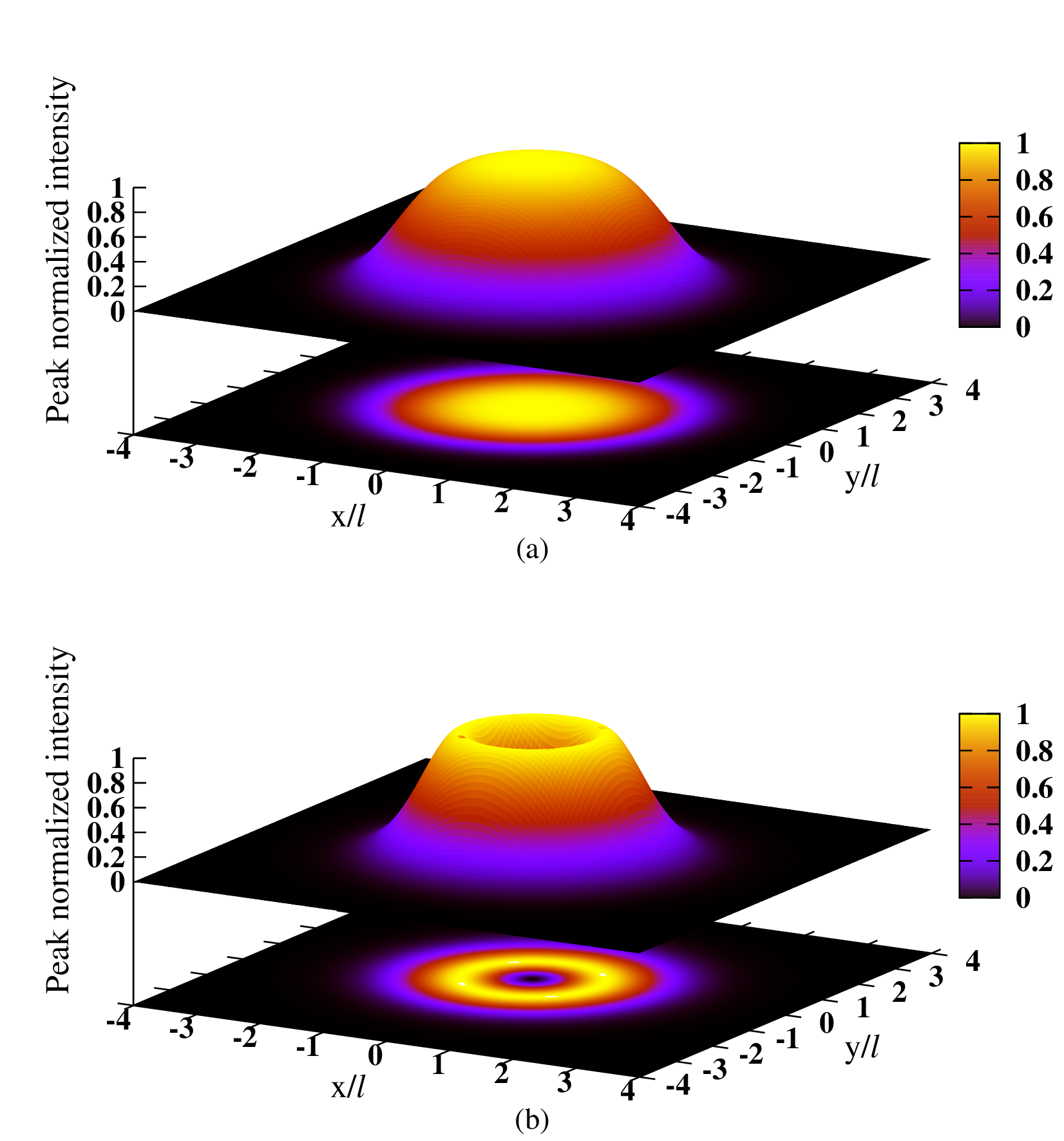}
	\caption{The 3-D intensity distribution of (a) SG beam of width 1.7$l$ and (b) LG spatiotemporal beam of width $l$}
	\label{fig:4}
\end{figure}
The next pulse is peaked at $\gamma_nt$ = 15,  Rabi frequency $\Omega_{D}^0 = 10.4\gamma_{n}$, with a negative chirp $\alpha_D$ = - 3.24 ps$^{-2}$ which returns back the population to the ground state. As demonstrated in Fig. \ref{fig:3}, the influence of phonons is negligible at low temperature. However, as the temperature increases, such as at 50K, the efficiency of population transfer becomes hindered due to phonon decoherence effects. We have taken the additional parameters for InGaAs/GaAs QDs which are used in \cite{ME2}. The phonon cutoff frequency $\omega_b$ = 2.2 ps$^{-1}$, and we choose $\gamma = \gamma^\prime$ = 1 $\mu$eV. The second pulse is implemented in a short interval of the first to make the population return efficiently so that the two pulses act like an on-off switch. 
\begin{figure}[ht]
	\centering
	\includegraphics[width=\linewidth]{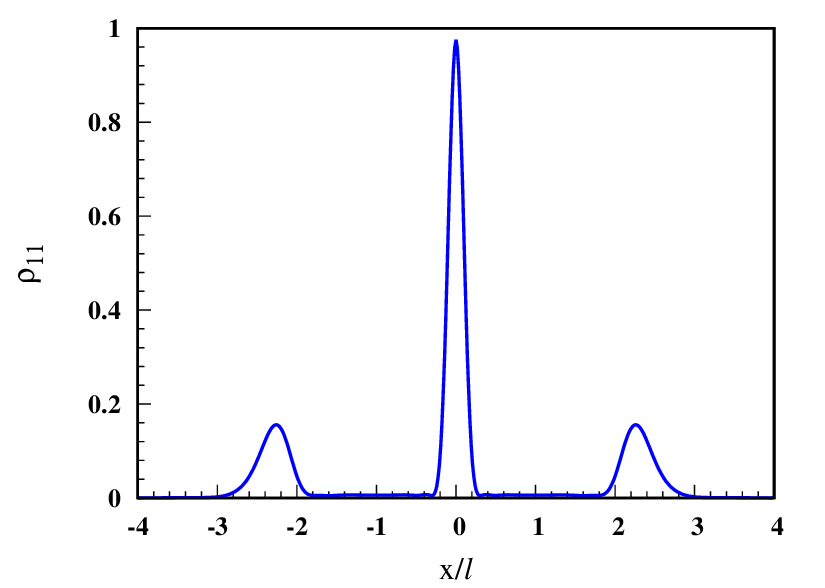}
	\caption{Population of the excited state vs. spatial extent in QD system. The applied spatiotemporal beams SG has waist $w_G$ = 1.7$l$, and LG has waist $w_D$ = $l$. The spatial distribution of the excited state population is shown at $\gamma_n$t = 30. Other parameters are the same as in Fig. \ref{fig:3}.}
	\label{fig:5}
\end{figure}
This efficient population transfer and return is crucial for imaging based on this scheme. The frequency-swept is adiabatic so that $\Omega_0^2 >> |d\Delta/dt|$.

Now for RAP-based imaging, the system is driven by two spatiotemporal beams with SG and LG spatial profiles shown in Figs. \ref{fig:4}(a) and \ref{fig:4}(b) respectively. The SG and LG beam waists are 1.7$l$  and 1.0$l$, respectively. Here $l$ is a characteristic length defined by $l$ = f/(k$\sigma$) \cite{intro11}. The lens f's focal length is 3.7mm., the wavevector $k = 2\pi n/\lambda$. We take the refractive index of the gallium phosphide (GaP) solid emersion lens, $n$ = 3.5, and wavelength of the laser $\lambda$ = 940 nm. The spatial extent of the beam before focusing on a lens is taken as $\sigma$ = 1.2 mm. So, the characteristic length $l$ becomes 131.86 nm. This can be regarded as the beam's spot size when it is focused through a lens for the above parameters. The Rabi frequency of the SG beam is $\Omega_{G}^0$ = 4.0$\gamma_n$ and for the LG beam, is $\Omega_{D}^0$ = 10.4$\gamma_n$. Successively implementing SG and LG spatiotemporal beams produce a spot size of  $\Delta \text{x}_{\text{FWHM}}/l$ = 0.2 $\textit{i.e.}$, approximately 26 nm at T = 4K. In Fig. (\ref{fig:5}), we show the excited state population distribution at $\gamma_n t$ = 30. 
\begin{figure}[ht]
	\centering
	\includegraphics[width=\linewidth]{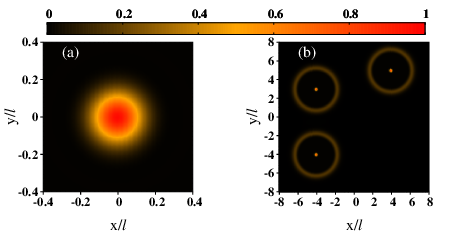}
	\caption{The 2-D spot size of (a) single QD emitter at (0,0). The outer ring is at a higher radius which is not shown here, (b) Multiple emitters at a preassigned position (-4,-4) bottom left, (-4,3) top left and (4,5) top right. Here we can see the low intense outer rings. At low temperature as phonon effect is negligible its effect is not considered. The parameters are the same as in Fig. \ref{fig:5}. }
	\label{fig:6}
\end{figure}
We observe that the central peak is not much distorted at lower temperatures. Other than the central peak, small side peaks appear due to the population transition to the excited state due to the LG beam. At higher temperature T = 50K we find the side peaks as well as the central peak distort considerably. The tail of the SG beam fails to take all the population to the excited state, leaving it partially in the ground state. However, the LG beam partially takes this leftover population to an excited state. We choose the beam waist judiciously so that the side peaks remain minimum and far apart from the central maximum. Figure \ref{fig:6}(a) shows the central spot in the 2d plane. Also, we can detect multiple QDs at different preassigned locations given in Fig. \ref{fig:6}(b). Thus, applying the SG and LG beams successively give a smaller spot size of the QDs emitters.
\begin{figure}[ht]
	\centering
	\includegraphics[width=\linewidth]{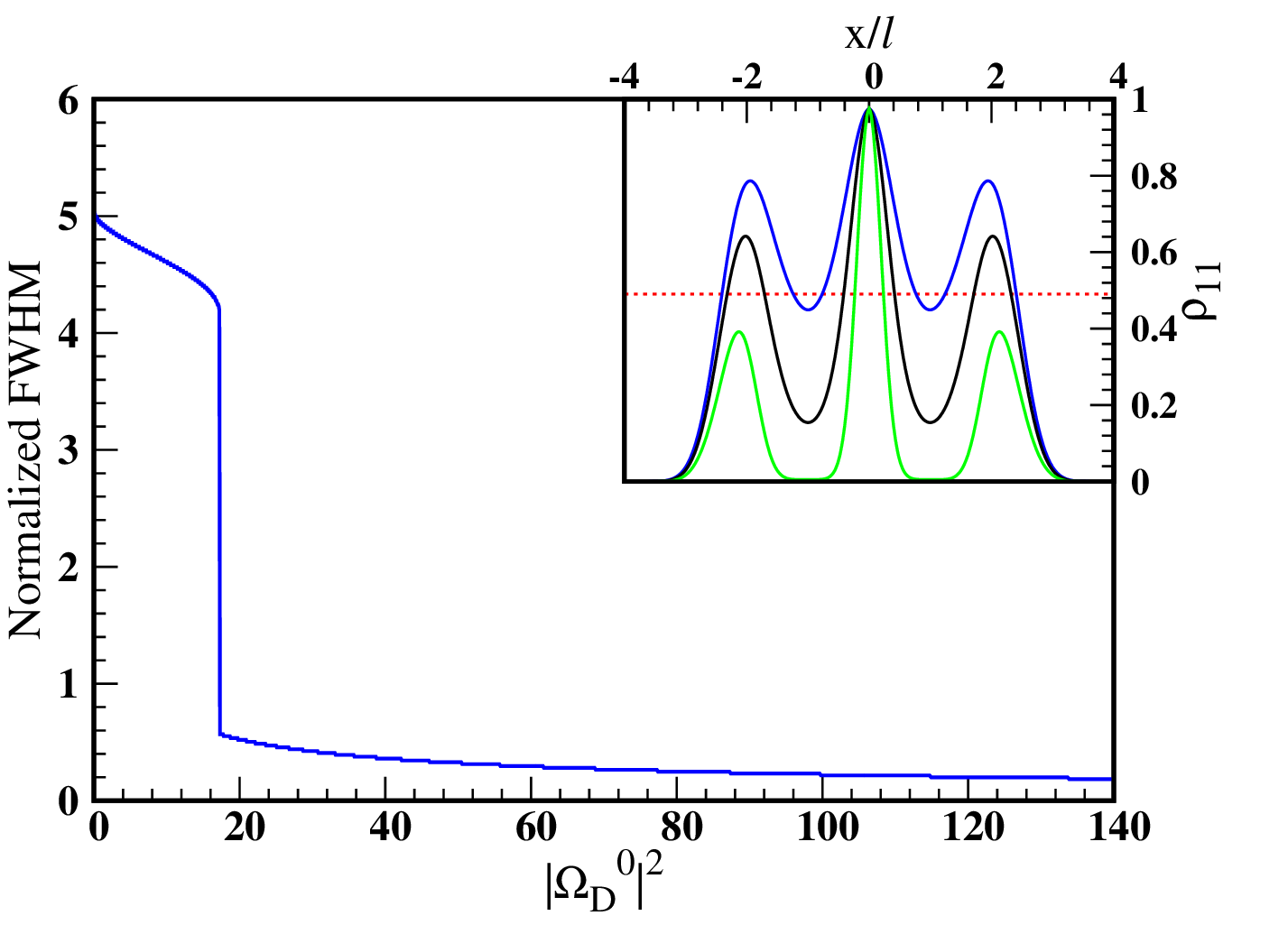}
	\caption{FWHM of the spot is plotted against the intensity of the spatiotemporal LG beam. The inset shows the excited state population for different LG intensities. The blue, black, and green solid lines correspond to $|\Omega_{D}^0|^2$ = 4, 9, 25 respectively. The red dotted line represents half of the central maximum.}
	\label{fig:7}
\end{figure}
As we observe at lower temperature the phonon effect is negligible on spot formation, we next analyzed the dependence of the FWHM of the QD spot with the intensity of the LG beam in Fig. \ref{fig:7}. The flat top portion of the SG beam, which exists between the spatial range $x/l=\pm1$, takes all population from the ground state to the excited state. The central peak develops due to the population presence in the excited state at the center $x/l=0$, where the intensity of the LG beam becomes zero. This is true for taking the orbital angular momentum index as 1 ($LG_0^{1}$) in Eq. (\ref{eq:LG}). This point singularity is the key to achieving sharp resolution at the QD center. The intensity peaks of the LG beam appear at $x/l=\pm1$, leading to stimulated emission from the excited state to the ground state. Subsequently, a dip is observed in the excited state population $\rho_{11}$ as shown in the inset of Fig. \ref{fig:7}. The central peak accompanied by shallow dips gives rise to a larger FWHM. By increasing the LG beam's intensity, the spatial distribution of the excited state population at $x/l=\pm1$ goes to zero very sharply. Hence, the inset of  Fig. \ref{fig:7} reveals the reason behind the narrow spot size formation, which can be made possible by the doughnut beam-assisted complete excited state depopulation. Before the sharp fall of the FWHM the doughnut beam intensity is very low. By slowly increasing the doughnut beam intensity and keeping the SG beam fixed, much of the population transfers to the ground state. In this context, the peak after the dip in the excited state population occurs due to tail of the LG beam where the intensity is lower than $x/l=\pm1$. The inset of Fig. \ref{fig:7} (blue and black solid lines) illustrates these phenomena. The red dotted line in the inset of Fig. \ref{fig:7} indicates the intensity level at half of the central maximum, while the green solid line illustrates a typical scenario in which a side peak occurs below this intensity threshold. It is to be noted that arbitrary enhancement of the intensity up to a larger extent is prohibited because of photobleaching or may even damage the live sample \cite{photobleaching1,photobleaching2}. 

{\color{black}
\subsection{Reduction of side peak}
It is evident from the previous analysis that the formation of side peaks around the central maximum of the QD spot is inevitable by considering the fields whose spatial envelopes represent by Eqs. (\ref{eq:field1}) and (\ref{eq:field2}). Even though these side peaks are small, these peaks may degrade the resolution where several QDs emitters are present within the specific range. So, finding the modulated fields which will decrease the side peak nearly to zero is pertinent. 
\begin{figure}[ht]
	\centering
	\includegraphics[width=\linewidth]{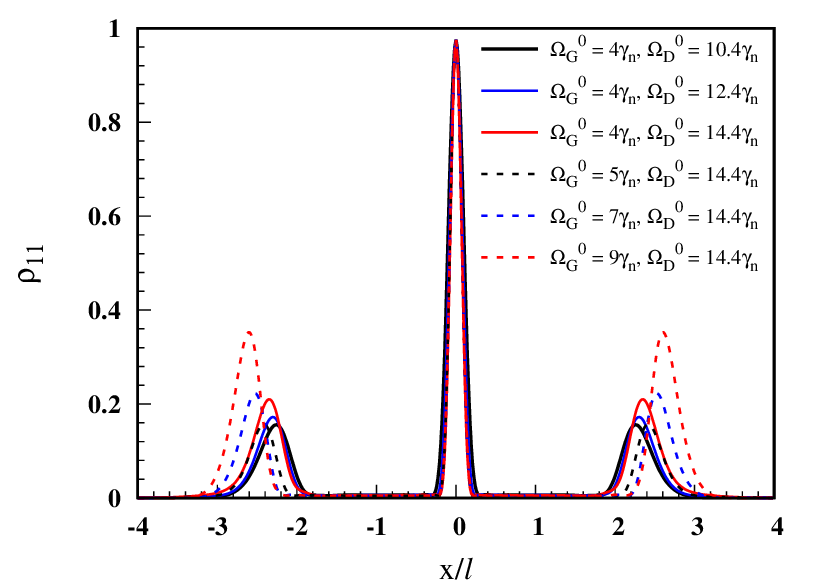}
	\caption{Normalized peak intensities of Bessel-modulated SG, Bessel-modulated LG beam and the population of the excited state are plotted against spatial extent. The red solid line, and the black dotted line correspond to the modulated LG and SG beams, respectively. Both the beams are truncated at $x/l = 1.05$. Other parameters are the same as in Fig. \ref{fig:4}.}
	\label{fig:8}
\end{figure}
A complete suppression of residual ground state population flow is possible by considering the truncated beams which are formed by the  Bessel-modulated SG and Bessel-modulated LG beams given as
\begin{subequations}
\begin{align}
	&\Vec{E}_{GM}(r,t) = \hat{\sigma}_-\mathcal{E}_{G}(r,t)\left[a_1 J_0(r^{36}) + a_2 J_2(r^{36})\right],\label{eq:mod_field1}\\
	&\Vec{E}_{DM}(r,t) = \hat{\sigma}_-\mathcal{E}_{D}(r,t)\left[b_1 J_0(r^{36}) + b_2 J_2(r^{36})\right]. \label{eq:mod_field2}
\end{align}
\end{subequations}
Here $J_i (i\in 0, 2)$ is the Bessel function of the first kind with order $i$, $a_i$ and $b_i$ are modulation coefficients (real numbers) which we have chosen, $a_1 = a_2 = 1.5$, $b_1 = b_2 = 1.0$. The modified Rabi frequencies of modulated LG and SG beams are $\Omega_{DM}$, $\Omega_{GM}$ respectively. In Fig. \ref{fig:8}, the two modulated beams are truncated at $x/l = \pm 1.05 $, which can be obtained experimentally by finite apertures. The modulated SG beam takes all the population from the ground state to the excited state. Due to the sharp fall of the modulated SG intensity at $x/l = \pm 1.05 $, the ground state population beyond the spatial range, $x/l = 0$ to $x/l = \pm 1.05$ cannot be excited to the excited state. Similarly, the excited state population goes through stimulated emission by the modulated LG beam keeping only the population at the center $x/l = 0$. Keeping the spot size of the QD emitter the same as Fig. \ref{fig:5}, we can reduce the side peak nearly to zero. The experimental realizations have demonstrated that by coupling $\sigma^-$ polarized light to a diamond-like QD system, the multilevel structure can effectively be reduced to a two-level system, which makes coherent excitation schemes experimentally feasible \cite{intro11}. In our work, the system parameters are chosen based on values reported in the corresponding experimental studies to ensure that our theoretical model remains experimentally realistic. Furthermore, in our scheme the unwanted side lobes in the excitation profile are suppressed by employing two structured beams. Such structured beam profiles can be generated using well-established optical beam-shaping techniques. For example, Bessel-like beams can be produced using an axicon (conical lens) \cite{reduction_1}, while more complex spatial modes such as LG beams or tailored SG profiles can be generated using a spatial light modulator (SLM) or by employing an annular aperture \cite{reduction_2}. These optical elements are routinely used in modern photonics and microscopy experiments, making the proposed excitation scheme experimentally accessible.
\subsection{Decoupling of phonons}
The phonon-induced decay rates will affect the population distribution among the QD states for different temperatures. To study the dependence we have solved the variational ME in Eq. (\ref{eq:variational master equation}) where all the decoherence rates are taken into account. The variational ME approach provides a unified and consistent framework for analyzing systems under both strong driving and strong exciton–phonon interactions. Notably, it seamlessly incorporates the weak-coupling and polaron ME as its limiting cases. 
 \begin{figure}[ht]
	\centering
	\includegraphics[width=\linewidth]{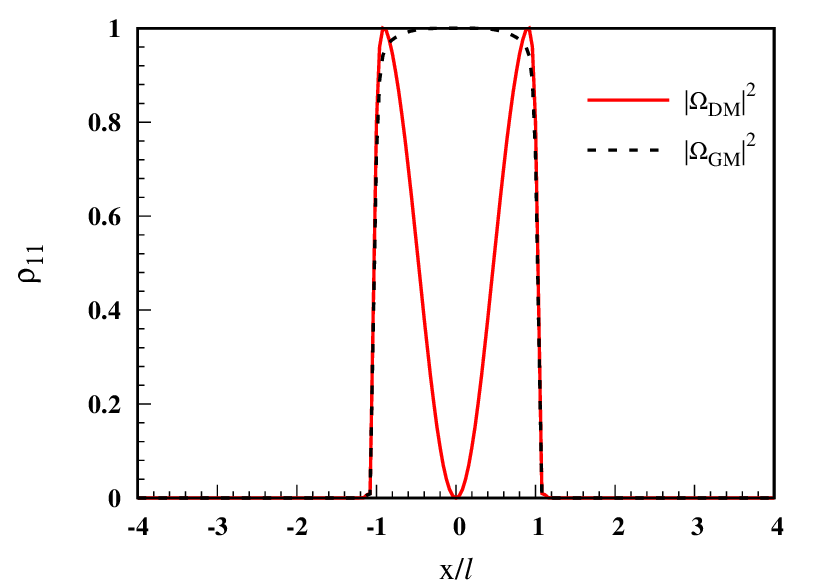}
	\caption{QD excited state population as a function of pulse area. (a) Transform-limited (pulse without chirping) Gaussian and (b) Gaussian chirp pulse of chirping 3.24 ps$^{-2}$ and a pulse of width 1.3$\tau_n$ is used. The electron-phonon coupling strength $\alpha_p$ = 0.027 ps$^2$.}
	\label{fig:9}
\end{figure}
A Gaussian pulse interacting with a two-level QD system induces Rabi oscillations (RO) of the excited state population. In Fig. \ref{fig:9}(a), we present the excited
state population as a function of the pulse area of a transform-limited (pulse without chirping) Gaussian pulse. Typically, RO in quantum dots are measured by recording the rotation of the Bloch vector after a pulse of specified duration as a function of the pulse area. In the literature \cite{intro21}, these signals are
\begin{figure}[ht]
	\centering
	\includegraphics[width=\linewidth]{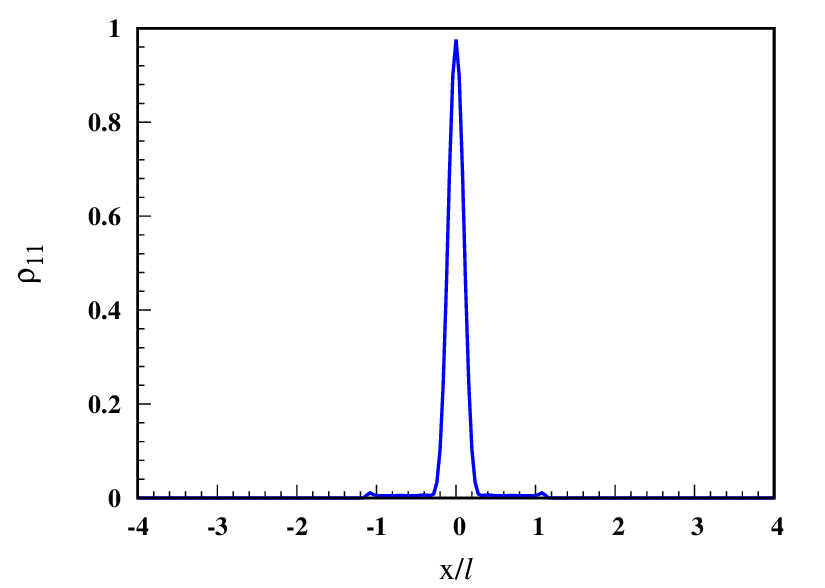}
	\caption{Spot size of QD emitter including and without including phonon coupling. Small area $\Omega_{SG}^0 = 8\gamma_n$, $\Omega_{D}^0$ = 20.8$\gamma_n$. Larger $\Omega_{SG}^0 = 12\gamma_n$, $\Omega_{D}^0$ = 31.2$\gamma_n$. Chirping and other parameters remain same as Fig. \ref{fig:9}. The modulation coefficients for truncated beams are $a_1 = a_2 = 1.5$, $b_1 = b_2 = 1.0$. A spot size of  $\Delta \text{x}_{\text{FWHM}}/\ell$ = 0.08 $\textit{i.e.}$, approximately 10 nm is observed at T = 10K.}
	\label{fig:10}
\end{figure}
characterized as Rabi rotations (RR) to differentiate them from the temporal evolution occurring during the pulse, which is referred to as RO. In Fig. \ref{fig:9}(a), the Gaussian pulse is centered at $\gamma_n t = 0$. Our findings indicate that the damping of RR displays a non-monotonic dependence on the pulse area. For small pulse areas, the amplitude of the RR diminishes with increasing pulse area, reflecting the expected damping behavior due to the interaction between a QD exciton and acoustic phonons. Conversely, for large pulse areas, the amplitude increases again, an observation that resembles undamping. This behavior is thus
termed the ‘reappearance of RR’. It is important to note that the reappearance of RR differs from the collapse-and-revival phenomenon \cite{jaynes} associated with the time-dependent RO in the well-known Jaynes-Cummings (JC) model, which exhibits periodic collapse and revival of RO in the time domain. In contrast, the reappearance of RR as a function of pulse area does not experience subsequent decay. We also note that at T = 4K and T = 50K, the amplitude of the RR is higher and lower, respectively, due to temperature-dependent phonon effects. At small pulse areas, the RR can be described using the polaron transformation; however, as the pulse area increases and satisfies the condition $|\Omega| > \omega_b$, the polaron ME no longer remains valid. At higher pulse areas, the variational theory can accurately describe the exciton-phonon decoupling effect observed in weak-coupling dynamics. The damping of the RR is also affected by the size of the quantum dot \cite{qdsize}. Furthermore, the reappearance of the RR is significantly dependent on the pulse width of the excitation \cite{qdsize2}. Utilizing positively chirped Gaussian pulse with short pulse width allows for the observation of the reappearance regime at reduced pulse areas. Figure \ref{fig:9}(b) illustrates the excited state population when a positively chirped Gaussian pulse is used. Our investigation indicates that a pulse area greater than $3\pi$ can efficiently transfer population to the excited state. Fulfilling the adiabatic condition, as discussed earlier, the population transfer becomes robust, unlike in transform-limited pulses, and is also less sensitive to intensity fluctuations. It is evident that the  distortion in QD spot could be reduced by decoupling exciton-phonon interactions when stronger spatiotemporal beams are employed. We plot the excited state population in Fig. \ref{fig:10} as a function of spatial extent. This shows that at T = 10K and with a small pulse area, the spatiotemporally modulated beam creates a spot that is subject to distortion. However, as the beam strength increases, a decoupling regime can be achieved where phonon effects are effectively mitigated. Additionally, as the temperature increases, there is a gradual reemergence of distortion. Significant distortion can be identified in the inset of Fig. \ref{fig:10}. At T = 10K, the side peaks remain below 6\% of the central maximum, which is negligible. This negligible spike originates from the abrupt truncation of the high-order Bessel-modulated beam profile at $x/l = \pm 1.05$. However, at higher temperatures, such as T = 50K, these side peaks exceed 10\% of the central maximum, degrading the resolution in an ensemble of quantum dots. Beyond T = 50 K, the central maximum also experiences significant distortion, severely compromising the sharpness of the spot. From Fig. 10, it is clear that, for a given temperature, we can decouple the phonon contribution by increasing the applied pulse area, thereby producing a tightly focused, smaller spot size. In this system, we have used a 940 nm driving light. After successfully applying this RAP-based STED technique, we achieve a spot-size resolution of $\Delta x_{FWHM}/l = 0.08$, corresponding to approximately 10 nm. In confocal microscopy, the theoretical resolution limit is $\lambda/2$, i.e., 470 nm for this system. In contrast, the current technique provides a resolution of approximately 10 nm, which is 47 times lower than the theoretical limit of conventional microscopy. Such a super-resolution imaging technique could have potential applications across domains such as neuroscience, live-cell imaging, materials science and nanomaterials, biomedical diagnostics, and drug delivery.
\section{CONCLUSION}
In conclusion, we have theoretically studied RAP-based imaging in semiconductor QD systems. For this purpose, we have adopted a semi-classical treatment for describing the spatiotemporal beams that interact with the two-level QD system. We use dressed state analysis to understand the RAP-assisted population transfer in the presence of chirping pulses. The phonon vibration in a semiconductor quantum dot system is inevitable at finite temperatures. To encompass the temperature-dependent phonon-induced decoherence rates for super-resolution imaging formation, we explore variational ME. We also show how the QD emitter's spot size depends on the LG field intensity. The compromise of image resolution in an ensemble of dense QDs is because of the circular ring. The dominant character of the LG beam tail over the SG beam tail causes residual population flow from the ground state to the excited state, creating the circular ring. Further, we have used the modulated truncated beams to overcome the image distortion.  Our analysis indicates that, although phonon-induced decoherences at finite temperatures adversely affect image resolution, an increase in field strength allows for the mitigation of these decoherences. This can be achieved by entering a regime where the exciton-phonon interaction can be effectively decoupled. Hence, this investigation may have potential applications in nano-scale imaging, with scalability and controllability. 
\begin{acknowledgments}
	We gratefully acknowledge funding by the Department of Science and Technology, Anusandhan National Research Foundation, Government of India (Grant No. CRG/2023/001318).
\end{acknowledgments}

\newpage
\onecolumngrid
\vspace{1cm}
\appendix*
\section{Derivation of phonon induced decay rates in variational ME}
In this Appendix, we give the derivation of analytically obtained scattering rates for the phonon-mediated scattering processes.
The variational ME is,
\begin{equation}
\frac{\partial \rho }{\partial t}=-\frac{i}{\hbar}\left[ H_{SV}(t),\rho  (t)\right]+\frac{\gamma}{2} \mathcal{L}\left[\sigma ^-\right]\rho +  \dfrac{ \gamma^\prime}{2}\mathcal{L}\left[\sigma ^+ \sigma ^-\right]\rho + \mathcal{L}_{ph}\rho,
\end{equation}
where,  $\mathcal{L}_{ph}$ is given by,
\begin{equation}
	\label{eq:A.2}
	\mathcal{L}_{ph}\rho = -\dfrac{1}{\hbar^2}\int_{0}^{\infty}\sum_{ij}\ d\tau \ \{ C_{ij}(\tau)[X_i(t),X_j(t,\tau)\rho(t)] + H.c. \}. \quad (i,j\in x,y,z)
\end{equation}
Here, $X_j(t,\tau) = e^{-i H_{SV}(t)\tau /\hbar} X_j(t) e^{i H_{SV}(t)\tau /\hbar}$. Now, the modified system operators are given below. Note that the time dependent field with Rabi frequency, $\Omega(t) = \Omega_D + \Omega_G e^{i\delta t}$ couples with QD system.
\begin{align}
	X_x(t) &= \dfrac{\hbar}{2}(\Omega(t)\sigma^+ + \Omega(t)^{*} \sigma^-)\nonumber\\ 
	&=\dfrac{\hbar}{2}\{\text{Re}[\Omega(t)]\sigma_x - \text{Im}[\Omega(t)]\sigma_y\}\\
	X_y(t) &= \dfrac{i\hbar}{2}(\Omega(t)\sigma^+ - \Omega(t)^{*} \sigma^-)\nonumber\\
	&=-\dfrac{\hbar}{2}\{\text{Im}[\Omega(t)]\sigma_x + \text{Re}[\Omega(t)]\sigma_y\}.\\
	X_z(t) &= \dfrac{\hbar}{2}(I+\sigma_z)
\end{align}
The correlation function has already been defined in Eqs. (\ref{polaron_correlation}) and Eqs. (\ref{weak_cross_correlation}).
The variational system Hamiltonian $H_{SV} = -\hbar \Delta_v |1 \rangle \langle 1| + \langle B \rangle X_x(t)$. Using Born-Markov approximation, the two-time phonon system operators can be written in terms of the one-time operators in the variational picture as
\begin{align}
	X_x(t, \tau) &=  e^{-i H_{SV}(t)\tau /\hbar} X_x(t) e^{i H_{SV}(t)\tau /\hbar},\\
	X_y(t, \tau) &=  e^{-i H_{SV}(t)\tau /\hbar} X_y(t) e^{i H_{SV}(t)\tau /\hbar},\\
	X_z(t, \tau) &=  e^{-i H_{SV}(t)\tau /\hbar} X_z(t) e^{i H_{SV}(t)\tau /\hbar}
\end{align}
So, we get the following,
\begin{align}
X_x(t, \tau) =& \dfrac{\hbar}{2}\biggl\{\biggl[\dfrac{\text{Re}[\Omega(t)](\Delta_v^2\ \text{cos}[\eta(t)\tau] + |\Omega_{R}(t)|^2)}{\eta(t)^2}-\dfrac{\text{Im}[\Omega(t)](\Delta_v\ \text{sin}[\eta(t) \tau])}{\eta(t)}\biggl]\sigma_x- \dfrac{2\Delta_v\langle B \rangle |\Omega(t)|^2 (1-\cos[\eta(t)\tau])}{\eta(t)^2} \nonumber\\
&\times \sigma^+ \sigma^--\biggl[\dfrac{\text{Im}[\Omega(t)](\Delta_v^2\  \text{cos}[\eta(t)\tau] + |\Omega_{R}(t)|^2)}{\eta(t)^2}+\dfrac{\text{Re}[\Omega(t)](\Delta_v\ \text{sin}[\eta(t)\tau])}{\eta(t)}\biggl]\sigma_y \biggl\},  	
\end{align}

\begin{align}
	X_y(t, \tau) =& -\dfrac{\hbar}{2}\biggl\{\biggl[\dfrac{\text{Re}[\Omega(t)](\Delta_v\  \text{sin}[\eta(t)\tau])}{\eta(t)}+\text{Im}[\Omega(t)]\text{cos}[\eta(t)\tau]\biggl]\sigma_x +\biggl[\text{Re}[\Omega(t)]\text{cos}[\eta(t)\tau]-\dfrac{\text{Im}[\Omega(t)](\Delta_v\ \text{sin}[\eta(t)\tau])}{\eta(t)}\biggl] \nonumber\\
	&\times \sigma_y+ \dfrac{2\langle B \rangle |\Omega(t)|^2 \text{sin}[\eta(t)\tau]}{\eta(t)}\sigma^+ \sigma^-\biggl\},  	
\end{align}
\begin{align}
	X_z(t, \tau) =& \hbar \biggl\{ \biggl[1 - \dfrac{|\Omega_{R}(t)|^2}{\eta^2} + \dfrac{|\Omega_{R}|^2 \cos(\eta \tau)}{\eta^2}\biggr]\sigma^+\sigma^- + \biggl[\dfrac{\langle B \rangle \Delta_v \cos (\eta \tau)}{2\eta^2} - \dfrac{\langle B \rangle \Delta_v}{2\eta^2}\biggr](\Omega(t)\sigma^+ + \Omega(t)^* \sigma^-) - \dfrac{i\langle B \rangle \sin(\eta \tau)}{2\eta}\nonumber \\ \times &(\Omega^*\sigma^- -\Omega \sigma^+)\biggr\} 	
\end{align}
where $\eta(t)=\sqrt{\Omega_{R}(t)^2 + \Delta_v^2}$, and $\Omega_{R}(t)= \langle B \rangle |\Omega(t)|$. Now from Eqn (\ref{eq:A.2}) we get
\begin{align}
	\mathcal{L}_{ph}\rho =& -\dfrac{1}{\hbar^2}\int_{0}^{\infty}\{ C_{xx}(\tau)[X_x(t),X_x(t,\tau)\rho(t)] + \textit{H.c.} + C_{yy}(\tau)[X_y(t),X_y(t,\tau)\rho(t)] + \textit{H.c.} \nonumber \\ &+ C_{zz}(\tau)[X_z(t),X_z(t,\tau)\rho(t)] + \textit{H.c.} + C_{yz}(\tau)[X_z(t),X_z(t,\tau)\rho(t)] + \textit{H.c.}\}\ d\tau.
\end{align}
We first solve for polaron theory by putting the values of $X_x(t,\tau)$ and $X_y(t,\tau)$ and then for the weak-coupling and cross-coupling case considering the $X_z(t,\tau)$ term. The following definition of real parameters are used.
\begin{align}
	f(t,\tau) &=\dfrac{\Delta^2\ \text{cos}[\eta(t)\tau] + |\Omega_R(t)|^2}{\eta(t)^2},\\
	g(t,\tau) &=\dfrac{\Delta\ \text{sin}[\eta(t)\tau]}{\eta(t)},\\
	h(t,\tau) &=\dfrac{2\Delta\langle B \rangle |\Omega(t)|(1-\text{cos}[\eta(t)\tau])}{\eta(t)^2},\\
	q(t,\tau) &= \text{cos}[\eta(t)\tau],\\
	r(t,\tau) &=\dfrac{2\langle B \rangle |\Omega(t)|^2\ \text{sin}[\eta(t)\tau]}{\eta(t)},\\
	m(t,\tau) &= \dfrac{\eta(t)^2 - |\Omega_R|^2 + |\Omega_R|^2 \cos(\eta(t) \tau) }{\eta(t)^2}, \\
	 n(t,\tau) &= \dfrac{\langle B \rangle \Delta_v }{2\eta(t)^2}\left\{\cos(\eta(t)\tau) - 1\right\}, \\
	 u(t,\tau) &= \dfrac{\langle B \rangle }{2\eta(t)}\sin(\eta(t)\tau).
\end{align}
For the polaron theory we get 
\begin{align}
	\mathcal{L}_{ph}^{polaron}\rho =& -\dfrac{1}{\hbar^2}\int_{0}^{\infty}\{ C_{xx}(\tau)[X_x(t),X_x(t,\tau)\rho(t)] + \textit{H.c.} + C_{yy}(\tau)[X_y(t),X_y(t,\tau)\rho(t)] + \textit{H.c.}  \}d\tau\nonumber \\
	=& -\dfrac{1}{4}\int_{0}^{\infty}\biggl\{ G_g(\tau)\biggl[(\text{Re}[\Omega(t)]\sigma_x - \text{Im}[\Omega(t)]\sigma_y),\biggl((\text{Re}[\Omega(t)]f(t,\tau)-\text{Im}[\Omega(t)]g(t,\tau))\sigma_x - (\text{Im}[\Omega(t)]f(t,\tau)\nonumber\\
	&+\text{Re}[\Omega(t)]g(t,\tau))\sigma_y - h(t,\tau)\sigma^+ \sigma^-\biggl)\rho(t)\biggl]\ +\ \textit{H.c.} + G_u(\tau)\biggl[(\text{Im}[\Omega(t)]\sigma_x + \text{Re}[\Omega(t)]\sigma_y),-\biggl((\text{Re}[\Omega(t)]g(t,\tau) \nonumber\\
	&+\text{Im}[\Omega(t)]q(t,\tau))\sigma_x + (\text{Re}[\Omega(t)]q(t,\tau)-\text{Im}[\Omega(t)]g(t,\tau))\sigma_y+r(t,\tau)\sigma^+ \sigma^-\biggl)\rho(t)\biggl]\ +\ \textit{H.c.}\biggl\}\ d\tau.
\end{align}
\begin{align}
	\mathcal{L}_{ph}^{polaron}\rho=&\int_{0}^{\infty}\biggl\{\biggl[-\dfrac{|\Omega(t)|^2}{4}(-\text{Re}[G_g(\tau)]f(t,\tau)+\text{Im}[G_g(\tau)+G_u(\tau)]g(t,\tau)-\text{Re}[G_u(\tau)]q(t,\tau))\mathcal{L}[\sigma^+]\rho\biggl]+\biggl[-\dfrac{|\Omega(t)|^2}{4}\nonumber\\
	 &\times (-\text{Re}[G_g(\tau)] f(t,\tau)-\text{Im}[G_g(\tau)+G_u(\tau)]g(t,\tau)-\text{Re}[G_u(\tau)]q(t,\tau))\mathcal{L}[\sigma^-]\rho\biggl]-\biggl[\dfrac{1}{2}((\text{Re}[G_u](\tau)q(t,\tau)\nonumber\\&-\text{Re}[G_g(\tau)]f(t,\tau)) (\text{Re}[\Omega(t)^2]-\text{Im}[\Omega(t)^2])-2\text{Re}[\Omega(t)]\text{Im}[\Omega(t)](\text{Re}[G_u(\tau)]-\text{Re}[G_g(\tau)]g(t,\tau)))(\sigma^+ \rho \sigma^+ \nonumber\\&+ \sigma^- \rho \sigma^-)\biggl]-i\biggl[ \dfrac{1}{2}((\text{Re}[G_u(\tau)]-\text{Re}[G_g(\tau)])(\text{Re}[\Omega(t)^2]-\text{Im}[\Omega(t)^2])g(t,\tau)+2\text{Re}[\Omega(t)]\text{Im}[\Omega(t)](\text{Re}[G_u(\tau)]q(t,\tau) \nonumber\\&-\text{Re}[G_g(\tau)] f(t,\tau)))(\sigma^+ \rho \sigma^+- \sigma^- \rho \sigma^-)\biggl]+i\biggl[ \dfrac{|\Omega(t)|^2}{2}((\text{Re}[G_g(\tau)]+\text{Re}[G_u(\tau)])g(t,\tau))[\sigma^+ \sigma^-,\rho]\biggl]-\biggl[i\dfrac{1}{4}(\text{Im}[\Omega(t)]\nonumber\\&\times G_g(\tau)h(t,\tau)+\text{Re}[\Omega(t)]G_u(\tau) r(t,\tau))(\sigma^+ \sigma^- \rho \sigma^+ + \sigma^- \rho - \sigma^+ \sigma^- \rho \sigma^-) + \textit{H.c.} \biggl]-\biggl[\dfrac{1}{4}(\text{Re}[\Omega(t)]G_g(\tau)h(t,\tau) \nonumber\\&-\text{Im}[\Omega(t)]G_u(\tau)r(t,\tau)) (\sigma^+ \sigma^- \rho \sigma^+- \sigma^- \rho + \sigma^+ \sigma^- \rho \sigma^-) + \textit{H.c.} \biggl]\biggl\} d\tau,\\
	\mathcal{L}_{ph}^{polaron}\rho = & \dfrac{\Gamma^{\sigma^+}}{2}\mathcal{L}[\sigma^+]\rho + \dfrac{\Gamma^{\sigma^-}}{2}\mathcal{L}[\sigma^-]\rho - \Gamma^{cd}(\sigma^+ \rho \sigma^+ + \sigma^- \rho \sigma^-) - i \Gamma^{sd}(\sigma^+ \rho \sigma^+ - \sigma^- \rho \sigma^-) + i \Delta^{\sigma^+ \sigma^-}[\sigma^+ \sigma^-, \rho]\nonumber\\& - [i \Gamma_{gu+}(\sigma^+ \sigma^- \rho \sigma^+ + \sigma^- \rho - \sigma^+ \sigma^- \rho \sigma^-)+ \textit{H.c.}] - [\Gamma_{gu-}(\sigma^+ \sigma^- \rho \sigma^+ - \sigma^- \rho + \sigma^+ \sigma^- \rho \sigma^-) + \textit{H.c.}]\label{eq:A.22},
\end{align}
where all the phonon-mediated scattering rates which contributes to the polaron case are
\begin{align}
	\Gamma^{\sigma^+} = & -\dfrac{|\Omega(t)|^2}{2}\int_{0}^{\infty}(-\text{Re}[G_g(\tau)]f(t,\tau)+\text{Im}[G_g(\tau)+G_u(\tau)]g(t,\tau)-\text{Re}[G_u(\tau)]q(t,\tau))\ d\tau\nonumber\\
	=& \dfrac{|\Omega_R(t)|^2}{2}\int_{0}^{\infty}\biggl(\text{Re}\biggl\{\{\text{cosh}[\phi(\tau)]-1\}\biggl\{\dfrac{\Delta^2\ \text{cos}[\eta(t)\tau]+|\Omega_R(t)|^2}{\eta(t)^2}\biggl\}\ +\ \text{sinh}[\phi(\tau)]{\text{cos}[\eta(t)\tau]} \biggl\}\ -\ \text{Im}[e^{\phi(\tau)}-1]\nonumber\\&\biggl\{\dfrac{\Delta\ \text{sin}[\eta(t)\tau]}{\eta(t)} \biggl\}\biggl)\ d\tau,\\
	\Gamma^{\sigma^-} =& -\dfrac{|\Omega(t)|^2}{2}\int_{0}^{\infty}(-\text{Re}[G_g(\tau)]f(t,\tau)-\text{Im}[G_g(\tau)+G_u(\tau)]g(t,\tau)-\text{Re}[G_u(\tau)]q(t,\tau))\ d\tau\nonumber\\
	=& \dfrac{|\Omega_R(t)|^2}{2}\int_{0}^{\infty}\biggl(\text{Re}\biggl\{\{\text{cosh}[\phi(\tau)]-1\}\biggl\{\dfrac{\Delta^2\ \text{cos}[\eta(t)\tau]+|\Omega_R(t)|^2}{\eta(t)^2}\biggl\}\ +\ \text{sinh}[\phi(\tau)]{\text{cos}[\eta(t)\tau]} \biggl\}\ +\ \text{Im}[e^{\phi(\tau)}-1]\nonumber\\&\biggl\{\dfrac{\Delta\ \text{sin}[\eta(t)\tau]}{\eta(t)} \biggl\}\biggl)\ d\tau,\\
	\Gamma^{cd} =\ & \dfrac{1}{2}\int_{0}^{\infty}((\text{Re}[G_u(\tau)](\tau)q(t,\tau)-\text{Re}[G_g(\tau)]f(t,\tau))(\text{Re}[\Omega(t)^2]-\text{Im}[\Omega(t)^2])-2\text{Re}[\Omega(t)]\text{Im}[\Omega(t)](\text{Re}[G_u(\tau)]\nonumber\\&-\text{Re}[G_g(\tau)])g(t,\tau))\ d\tau \nonumber\\
	=\ &\dfrac{1}{2}\int_{0}^{\infty}\langle B \rangle^2 \biggl(\text{Re}\biggl\{\text{sinh}[\phi(\tau)]\text{cos}[\eta(t)\tau]-\{\text{cosh}[\phi(\tau)]-1\}\biggl\{\dfrac{\Delta^2\ \text{cos}[\eta(t)\tau]+|\Omega_R(t)|^2}{\eta(t)^2} \biggl\} \biggl\}(\text{Re}[\Omega(t)^2]\nonumber\\&-\text{Im}[\Omega(t)^2])-2\text{Re}[\Omega]\text{Im}[\Omega]\text{Re}[1-e^{-\phi(\tau)}]\biggl\{\dfrac{\Delta\ \text{sin}[\eta(t)\tau]}{\eta(t)} \biggl\} \biggl)\ d\tau,\\
	\Gamma^{sd}=\ &\dfrac{1}{2}\int_{0}^{\infty}((\text{Re}[G_u(\tau)]-\text{Re}[G_g(\tau)])g(t,\tau)(\text{Re}[\Omega(t)^2]-\text{Im}[\Omega(t)^2])+2\text{Re}[\Omega(t)]\text{Im}[\Omega(t)](\text{Re}[G_u(\tau)]q(t,\tau)\nonumber\\&-\text{Re}[G_g(\tau)]g(t,\tau)))\ d\tau \nonumber\\=\ &\dfrac{1}{2}\int_{0}^{\infty}\langle B \rangle^2 \biggl(\text{Re}[1-e^{-\phi(\tau)}]\biggl\{\dfrac{\Delta\ \text{sin}[\eta(t)\tau]}{\eta(t)} \biggl\}(\text{Re}[\Omega(t)^2]-\text{Im}[\Omega(t)^2])+2\text{Re}[\Omega(t)]\text{Im}[\Omega(t)]\text{Re}\biggl\{\text{sinh}[\phi(\tau)]\nonumber\\&\times \text{cos}[\eta(t)\tau]-\{\text{cosh}[\phi(\tau)]-1\}\biggl\{\dfrac{\Delta^2\ \text{cos}[\eta(t)\tau]+|\Omega_R|^2}{\eta(t)^2} \biggl\} \biggl\} \biggl)\ d\tau,\\
	\Delta^{\sigma^+ \sigma^-}=& \dfrac{|\Omega_R(t)|^2}{2}\int_{0}^{\infty}\text{Re}([G_g(\tau)]+G_u(\tau))g(t,\tau)\ d\tau\nonumber\\
	=&\dfrac{|\Omega_R(t)|^2}{2}\int_{0}^{\infty}\text{Re}[e^{\phi(\tau)}-1]\biggl(\dfrac{\Delta\ \text{sin}[\eta(t)\tau]}{\eta(t)} \biggl)\ d\tau,
\end{align}
\begin{align}
	\Gamma_{gu+} =\ & \dfrac{1}{4}\int_{0}^{\infty}(\text{Im}[\Omega(t)]G_g(\tau)h(t,\tau)+\text{Re}[\Omega(t)]G_u(\tau)r(t,\tau))\ d\tau\nonumber\\
	=\ & \dfrac{1}{4}\int_{0}^{\infty}\biggl(\text{Im}[\Omega(t)]\{\text{cosh}[\phi(t)\tau]-1\}\biggl\{\dfrac{2\Delta\langle B \rangle |\Omega(t)|^2(1-\text{cos}[\eta(t)\tau])}{\eta(t)^2} \biggl\}+\text{Re}[\Omega(t)]\text{sinh}[\phi(\tau)]\nonumber \\&\times\biggl\{\dfrac{2\langle B \rangle |\Omega(t)|^2 \text{sin}[\eta(t)\tau]}{\eta(t)} \biggl\} \biggl)\ d\tau,\\
	\Gamma_{gu-}=\ & \dfrac{1}{4}\int_{0}^{\infty}(\text{Re}[\Omega(t)]G_g(\tau)h(t,\tau)-\text{Im}[\Omega(t)]G_u(\tau)r(t,\tau))\ d\tau\nonumber\\
	=\ &\dfrac{1}{4}\int_{0}^{\infty}\biggl(\text{Re}[\Omega(t)]\{\text{cosh}[\phi(t)\tau]-1\}\biggl\{\dfrac{2\Delta\langle B \rangle |\Omega(t)|^2(1-\text{cos}[\eta(t)\tau])}{\eta(t)^2} \biggl\}-\text{Im}[\Omega(t)]\text{sinh}[\phi(\tau)]\nonumber\\&\times \biggl\{\dfrac{2\langle B \rangle |\Omega(t)|^2 \text{sin}[\eta(t)\tau]}{\eta(t)} \biggl\} \biggl)\ d\tau.
\end{align}
For the weak-coupling case the decay rates are
\begin{align}
	\mathcal{L}_{ph}^{weak} \rho =& -\int_{0}^{\infty} \biggl\{ \text{Re}[C_{zz}](\sigma^+\sigma^-\rho + \rho\sigma^+\sigma^- - 2\sigma^+\sigma^-\rho \sigma^+\sigma^-)m(t,\tau) + i\text{Im}[C_{zz}](\sigma^+\sigma^-\rho - \rho\sigma^+\sigma^-)m(t,\tau)\nonumber \\ & + \text{Re}[C_{zz}]\biggl[(\Omega\sigma^+\rho - \Omega\sigma^+\rho\sigma^+\sigma^- - \Omega^*\sigma^-\rho\sigma^+\sigma^-) + \textit{H.c.}\biggr]n(t,\tau) + i\text{Im}[C_{zz}]\biggl[(\Omega\sigma^+\rho - \Omega\sigma^+\rho\sigma^+\sigma^- \nonumber \\ &- \Omega^*\sigma^-\rho\sigma^+\sigma^-) - \textit{H.c.}\biggr]n(t,\tau) + i\text{Re}[C_{zz}]\biggl[\left(\Omega\sigma^+\rho - \Omega\sigma^+\rho\sigma^+\sigma^- + \Omega^*\sigma^-\rho\sigma^+\sigma^-\right) - \textit{H.c.}\biggr]u(t,\tau) \nonumber \\ &- \text{Im}[C_{zz}]\biggl[\left(\Omega\sigma^+\rho - \Omega\sigma^+\rho\sigma^+\sigma^- + \Omega^*\sigma^-\rho\sigma^+\sigma^-\right) + \textit{H.c.}\biggr]u(t,\tau)\biggr\} d\tau,  
\end{align}
and finally for the cross coupling we get
\begin{align}
	\mathcal{L}_{ph}^{cross} \rho =& -\dfrac{1}{2} \int_{0}^{\infty}\biggl\{i\text{Re}[C_{yz}]\biggl[\left(\Omega^*\sigma^-\rho - \Omega^*\sigma^+\sigma^-\rho\sigma^- + \Omega\sigma^+\sigma^-\rho\sigma^+\right) - \textit{H.c.}\biggr]m(t,\tau) - \text{Im}[C_{yz}]\biggl[(\Omega^*\sigma^-\rho \nonumber \\ &- \Omega^*\sigma^+\sigma^-\rho\sigma^- + \Omega\sigma^+\sigma^-\rho\sigma^+) + \textit{H.c.}\biggr]m(t,\tau) + i\text{Re}[C_{yz}]\biggl[\left(|\Omega|^2\sigma^-\sigma^+\rho - |\Omega|^2\sigma^+\sigma^-\rho\right) - \textit{H.c.}\biggr]n(t,\tau)\nonumber \\ & - \text{Im}[C_{yz}]\biggl[\left(|\Omega|^2\sigma^-\sigma^+\rho - |\Omega|^2\sigma^+\sigma^-\rho\right) + \textit{H.c.}\biggr]n(t,\tau) + i\text{Re}[C_{yz}]\left(-2(\Omega^*)^2\sigma^-\rho\sigma^- + 2\Omega^2\sigma^+\rho\sigma^+\right)n(t,\tau) \nonumber \\ &+ \text{Im}[C_{yz}]\left(2|\Omega|^2\sigma^+\rho\sigma^- - 2|\Omega|^2\sigma^-\rho\sigma^+\right)n(t,\tau) + \text{Re}[C_{yz}]\biggl[\left(-|\Omega|^2\sigma^-\sigma^+\rho - |\Omega|^2\sigma^+\sigma^-\rho\right) + \textit{H.c.}\biggr]u(t,\tau)\nonumber \\ & + i\text{Im}[C_{z}]\biggl[\left(-|\Omega|^2\sigma^-\sigma^+\rho - |\Omega|^2\sigma^+\sigma^-\rho\right) - \textit{H.c.}\biggr]u(t,\tau) + \text{Re}[C_{yz}](-2(\Omega^*)^2\sigma^-\rho\sigma^+ + 2|\Omega|^2\sigma^-\rho\sigma^+ \nonumber \\ &+ 2|\Omega|^2\sigma^+\rho\sigma^- - 2\Omega^2\sigma^+\rho\sigma^+)u(t,\tau)\biggr\} d\tau.  
\end{align}
The above phonon-induced scattering rates are function of time. At each time t the values are calculated by integrating with respect to $\tau$. These analytically derived scattering rates illustrate the physical picture of electron-phonon coupling in semiconductor QDs.
\newpage
\twocolumngrid
\bibliography{paper}
\end{document}